# Fast-neutron and gamma-ray imaging with a capillary liquid xenon converter coupled to a gaseous photomultiplier

I. Israelashvili[a,c], A. E. C. Coimbra[a], D. Vartsky[a], L. Arazi[b], S. Shchemelinin[a],

E. N. Caspi[c], A. Breskin[a]

[a] *Dept. of Astrophysics & Particle Physics, Weizmann Institute of Science, Rehovot, Israel*

[b] *Physics Core Facilities, Weizmann Institute of Science, Rehovot, Israel*

[c] *Physics Department, Nuclear Research Centre – Negev, Beer-Sheva, Israel*

Abstract

Gamma-ray and fast-neutron imaging was performed with a novel liquid xenon (LXe) scintillation detector read out by a Gaseous Photomultiplier (GPM). The 100 mm diameter detector prototype comprised a capillary-filled LXe converter/scintillator, coupled to a triple-THGEM imaging-GPM, with its first electrode coated by a CsI UV-photocathode, operated in Ne/5%CH$_4$ at cryogenic temperatures. Radiation localization in 2D was derived from scintillation-induced photoelectron avalanches, measured on the GPM's segmented anode. The localization properties of $^{60}$Co gamma-rays and a mixed fast-neutron/gamma-ray field from an AmBe neutron source were derived from irradiation of a Pb edge absorber. Spatial resolutions of 12±2 mm and 10±2 mm (FWHM) were reached with $^{60}$Co and AmBe sources, respectively. The experimental results are in good agreement with GEANT4 simulations. The calculated ultimate expected resolutions for our application-relevant 4.4 and 15.1 MeV gamma-rays and 1-15 MeV neutrons are 2-4 mm and ~2 mm (FWHM), respectively. These results indicate the potential applicability of the new detector concept to Fast-Neutron Resonance Radiography (FNRR) and Dual-Discrete-Energy Gamma Radiography (DDEGR) of large objects.

# 1 Introduction

Gamma-ray and fast-neutron imaging technologies are currently applied for investigating the content of aviation- and marine-cargo containers, trucks and nuclear waste containers (see for example [1, 2]). MeV-scale x-ray or gamma-ray radiographic inspection methods, such as Dual Energy Bremsstrahlung Radiography (DEBR) [3-5] or Dual-Discrete-Energy Gamma Radiography (DDEGR) [6], are used for the detection of concealed Special Nuclear Materials (SNM), providing high-resolution images of object shapes and densities and some selectivity between high-Z elements. DEBR makes use of continuous x-ray spectra, generated by accelerated electrons at two different bombarding energies. DDEGR relies on two discrete gamma-rays, of 4.4 MeV and 15.1 MeV, emitted by the $^{11}B(d,n\gamma)^{12}C$ reaction. Fast-neutron imaging methods, such as Fast-Neutron Resonance Radiography (FNRR) [7], utilize a broad neutron spectrum of 2-10 MeV to provide a sensitive probe for identifying low-Z elements such as H, C, N and O; these are the main constituents of explosives and narcotics. In addition, FNRR provides a means for identifying the type of the explosive by determination of the density ratios of its main constituent elements [7]. FNRR has been also proposed recently for determining of oil and water content in drilled formation cores [8].

The requirements from fast-neutron and gamma-ray detectors for contraband detections, are detection efficiency >10% and position resolution better than 5-10 mm for both types of radiation [9 and references therein].

A unique inspection system concept featuring both FNRR and DDEGR techniques could combine the capability of low-Z objects detection and elemental identification with high-Z objects selectivity. This requires intense radiation sources emitting fast neutrons and gamma-rays as well as an efficient imaging detector for both types of radiation. A suitable source is the one based on the $^{11}B(d,n\gamma)^{12}C$ reaction, with 3-7 MeV deuterons interacting with a thick $^{11}B$ target [6, 10]. In addition to the two discrete gamma rays (4.4 MeV and 15.1 MeV) the reaction yields a broad spectrum of fast neutrons; e.g., a 6 MeV deuteron beam yields an almost continuous neutron spectrum, with energies of up to ~18 MeV [6, 10].

This work is part of a feasibility study on a new cost-effective, large-area robust detector concept [11-13], for the simultaneous detection of gamma-rays and fast neutrons within the same detection medium. Gamma-ray spectroscopy is performed by pulse-height analysis, while fast-neutron spectroscopy and neutron/gamma discrimination is done by time-of-flight (TOF). Compared to imaging by two separate systems, such an approach would have practical advantages in terms of cost and throughput; it would also enable the use of the data without the need for geometrical alignments and corrections.

The proposed detector concept [11] investigated in this work comprises an efficient, fast liquid xenon (LXe) converter-scintillator contained within Tefzel [14] capillaries. Tefzel was selected because it is a hydrogen-rich polymer ($C_4F_4H_4$), with low refractive index compared to that of LXe at 178 nm ($n_{tefzel} = 1.5$ [15] vs. $n_{LXe} = 1.69$ [16]). The hydrogen content largely improves the position resolution for neutrons, due to a more efficient collisional energy transfer, which reduces the probability of a second neutron scattering vertex far from the first interaction point. Based on simulations [12], this effect should result in ~3-fold better position resolution compared to a plain LXe converter. The converter is coupled through a window to a UV-sensitive gaseous imaging photomultiplier (GPM) [17]. The GPM incorporates a cascaded structure of Thick Gas Electron Multiplier (THGEM) electrodes [18-21] with a CsI UV-sensitive photocathode deposited on the first electrode. Recent studies [22, 23] on a triple-THGEM GPM, have shown long-term stable operation at cryogenic temperatures (190-220 K), under UV irradiation and with alpha particles (using an immersed $^{241}$Am source); avalanche gains $> 10^5$ were reached in Ne/5%CH$_4$ and Ne/20%CH$_4$ allowing for the detection of photon pulses over a broad dynamic range (from one photoelectron to several thousand photoelectrons).

In this article we investigate the imaging performance of a 100 mm diameter active-area prototype detector using a $^{60}$Co gamma-ray source (1.17 and 1.33 MeV) and an AmBe source (0-11 MeV neutrons and 4.4 MeV gamma-rays). The results were validated by GEANT4 simulations, which were also extended to predict the performance of a projected operational large-area detector at higher application-relevant gamma-ray energies (up to 15.1 MeV) and 2-15 MeV neutrons - foreseen in FNRR and DDEGR.

## 2 Experimental setup and methodology

## 2.1 Experimental setup

### 2.1.1 *LXe cryostat*

The experiments were conducted using the Weizmann Institute Liquid Xenon cryostat (WILiX), described in detail elsewhere [24] (see Figure 1A). The inner vacuum chamber (IVC), modified with respect to [23] to meet the requirements of the present work, contained a radiation converter-scintillator consisting of ~5500 Tefzel [14] capillaries (outer diameter OD = 1.6 mm, inner diameter ID = 1.0 mm, length = 70 mm, see Figure 1B), filled with LXe. The overall diameter of the converter was 133 mm. The LXe level was set above the bottom face of the GPM UV-window. The distance from the top of the capillaries to the bottom face of the window was 6 mm. LXe was continuously extracted, purified and re-liquefied, as described in details in [23]. Under steady-state conditions the LXe temperature was ~170 K, at a pressure of 1.3 bar.

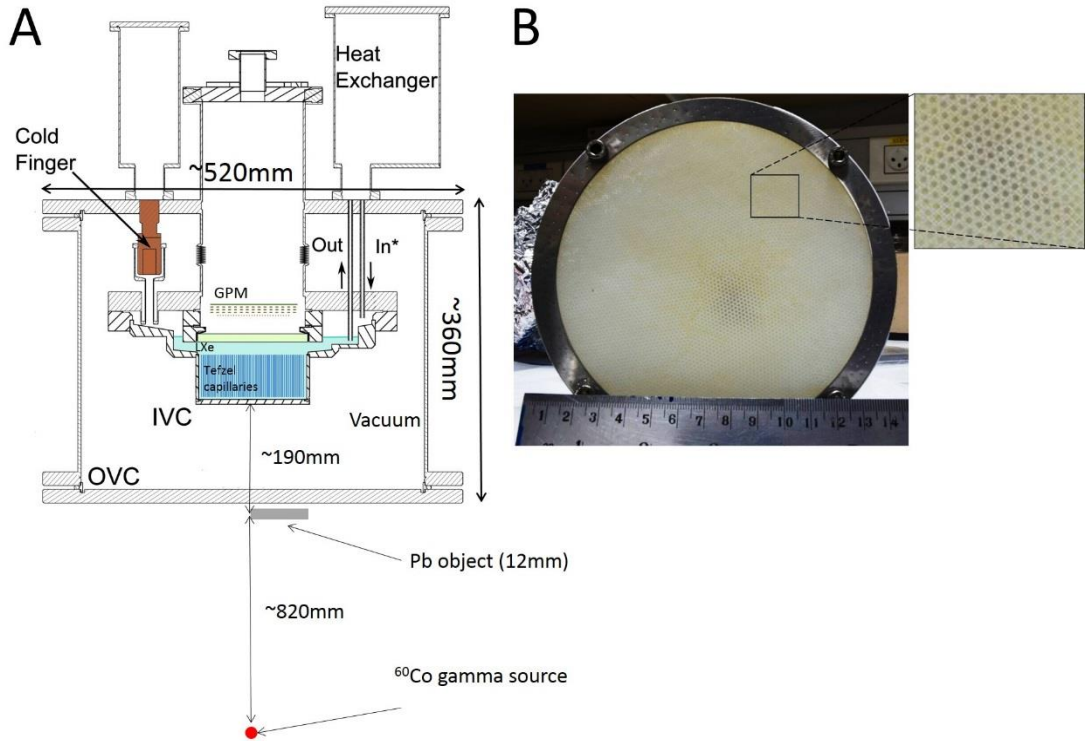

**Figure 1:** (A) The WILiX cryostat [23], with the GPM assembly operating with Ne/5%CH₄ and the Tefzel capillaries immersed in LXe, viewed by the GPM through a DUV-grade fused silica viewport. Imaging experiments were performed either with external collimated radiation sources or with a broad beam irradiating an object placed under the cryostat's bottom flange; shown is the configuration for gamma-ray imaging of a Pb object edge; see text for detailed gamma- and gamma/neutron-irradiation configurations. (B) The radiation converter composed of ~5500 Tefzel capillaries (OD = 1.6 mm, ID = 1.0 mm, length = 70 mm) assembled in a Teflon holder, between two meshes.

### 2.1.2  *Cryogenic GPM*

The GPM setup, shown in Figure 2, consisted of a cascaded structure of three THGEM electrodes, with a CsI photocathode deposited on the first, followed by a segmented readout electrode comprising 61 hexagonal pads (see layout and pad dimensions in Figure 3). The GPM's photocathode was located ~20 mm above the top surface of the window (34 mm above the top edge of the capillaries), viewing it through a DUV-grade fused silica viewport (clear diameter 137 mm, MPF part number: A0650-7-CF). The 0.4 mm-thick THGEM electrodes, Cu-clad and Au-plated on both sides (produced by ELTOS SpA, Italy, and further processed at the CERN PCB workshop) were made of FR4 with an active (perforated) diameter of 100 mm; the holes were arranged in a hexagonal pattern, with a hole diameter d = 0.4 mm, pitch a = 0.8 mm (between hole centers) and an etched hole rim h = 50 μm. The Cu-layer thickness (after etching) was ~64 μm. The transfer gaps between the stages, as well as the induction gap between the last THGEM and the segmented readout electrode were 1.5 mm wide. Each of the THGEM faces, as well as the mesh mounted 4.8 mm from THGEM1, had separate HV bias, provided through low-pass filters by CAEN N471A HV power supplies.

The GPM was operated along this study with Ne/5%CH$_4$ under a gas flow of 20 sccm, at pressures ranging from 475 to 647 mbar and a typical temperature of 210 K. The relatively low operating pressure, as well as the low CH$_4$ concentration (5%), were chosen to allow for better operation stability compared to the conditions used in [22] (475 mbar of Ne/20%CH$_4$). We attribute this change in stable conditions to the accumulation of discharge history on the particular THGEM electrodes used for the same studies.

The initial quantum efficiency (QE) value of the particular CsI photocathode used over this study was measured to be 22% at 175 nm before transferring it from the evaporation system to the GPM; the value re-measured after seven months of operation was 8%. The degradation in QE was likely the result of water outgassing from the top part of the GPM chamber, kept at room temperature throughout the experiments (while the photocathode temperature was ~210 K). It is not known if the QE degraded gradually along the 7-month operation, or over a short period of time.

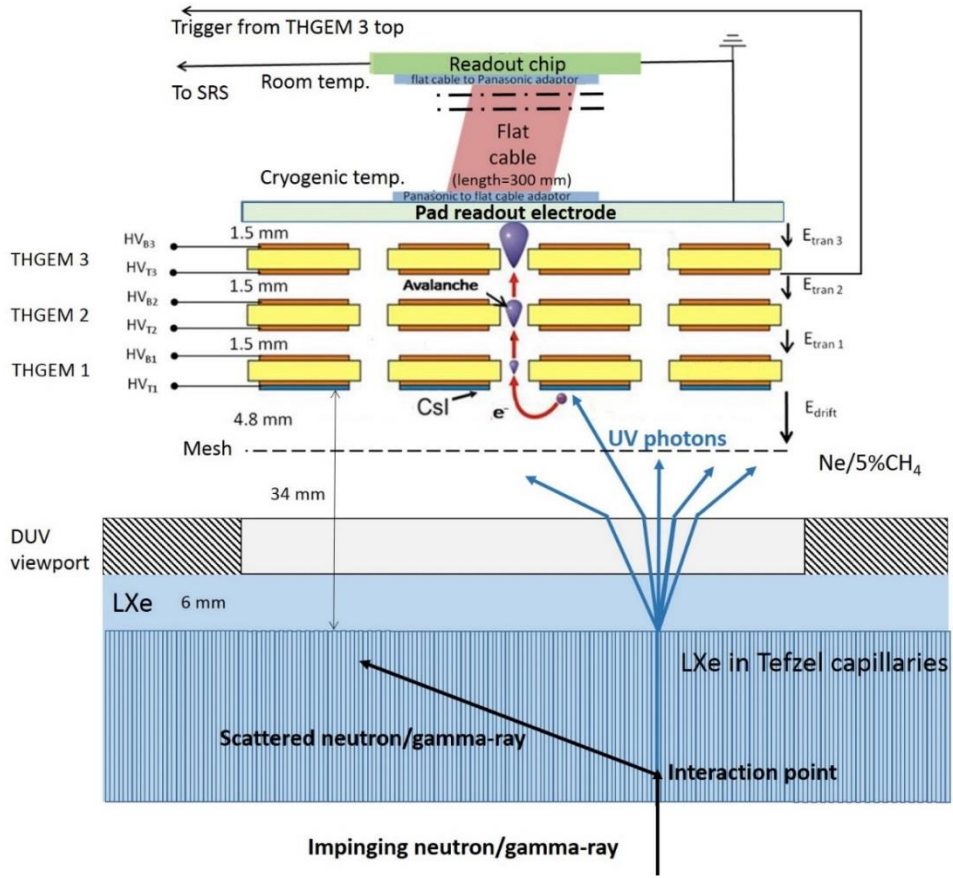

**Figure 2:** Schematic view of the GPM setup; a cascade of 3 THGEM electrodes, the first one coated with a reflective CsI photocathode, followed by a 2D readout pad electrode. Signals from individual pads are transmitted through a 300 mm long flat cable into a readout chip (kept at room temperature) and processed with SRS electronics [25] (see text).

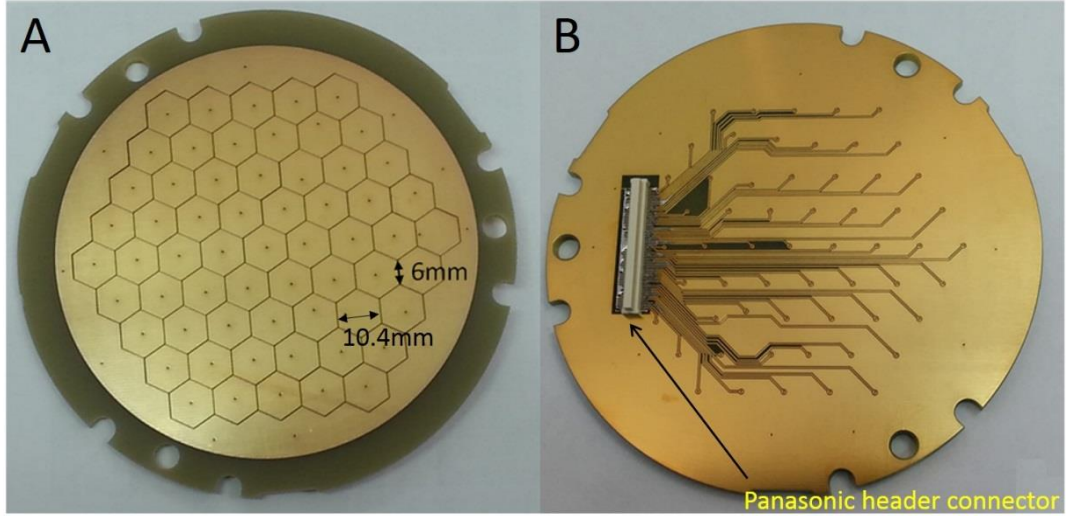

**Figure 3:** Segmented 61 hexagonal-pads readout electrode, front side (A) and back side (B). Pad side is 6 mm and its width is 10.4 mm. The gap between neighboring pads is 0.2 mm.

### *2.1.3   Readout electronics*

The segmented readout electrode, shown in Figure 3A, had 61 hexagonal pads (of 6 mm side and 10.4 mm width). Each of the pads was connected to an individual channel of the front-end analog hybrid readout chip via a Panasonic header connector (type: AXK6SA3677YG) (see Figure 3B) mounted on the pad electrode. The readout chip was not designed for cryogenic temperature operation (~200 K); therefore, the pad signals were transferred to a remotely-placed chip through a 30 cm long ribbon flat cable (3754/80 80 conductor 0.64 mm pitch) using two dedicated PCB adapters. The flat cable (placed inside the GPM gas vessel) was wrapped with a thin Cu ground-shielding foil. The readout chip was connected to the external SRS system [25] via a 1 m-long homemade vacuum-rated micro-HDMI-to-HDMI cable and feedthrough. Triggers for the SRS system were extracted from the THGEM3 electrode (see Figure 2) through a coaxial cable into a Canberra 2006 charge sensitive preamplifier located outside the GPM chamber. These trigger signals were shaped by a timing filter amplifier (Ortec model 474) followed by leading edge discriminator (Philips Scientific model 730) and then fed into the SRS trigger input.

Charge signals from the LXe/GPM detector were processed and saved by the SRS electronics, event-by-event, followed by offline analysis. Typical noise- and gamma-induced charge spectra in the detector, in each of the 61 readout pads, are shown in Figure 4. One can set a threshold on the charge ("charge threshold") and check, event-by-event, the number of pads exceeding threshold. In a similar way one can set a threshold on the number of firing pads ("pad threshold"), and for example exclude events with lower number of pads (for improving position resolution, but at the cost of losing some detection efficiency).

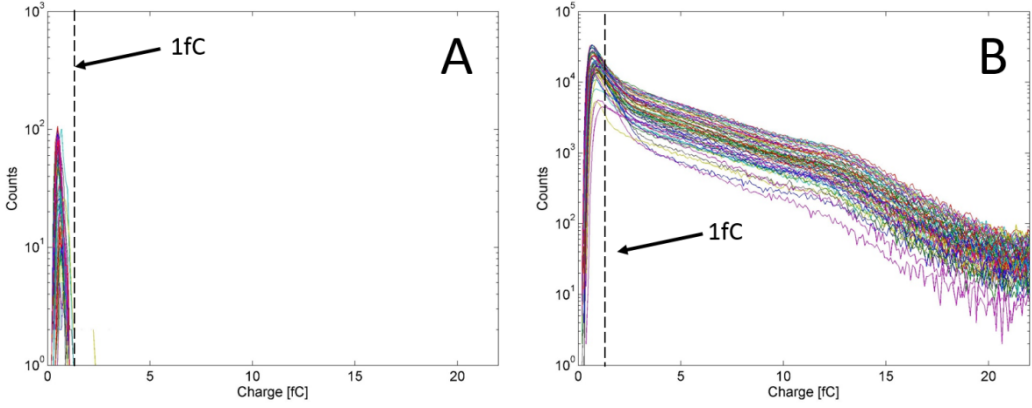

**Figure 4:** Charge spectra measured in each of the 61 readout pads. (A) Electronic noise and (B) $^{60}$Co gamma-rays detected by the LXe/GPM detector. The GPM gain in these measurements was $4\times10^4$.

For each event, the center of gravity (COG) of the firing pads is calculated according to equation 1 and a 2D histogram of the COG values can be plotted. In equation 1, $Q_{i,j}$ is the charge collected in pad j in event i and $\vec{P_j}$ is a [x, y] position vector of the center of pad j. Note that in cases where the number of PEs in each pad is small ($\lesssim 3$), calculating the simple unweighted COG, by setting $Q_{i,j} = 1$ for all firing pads (and zero for non-firing ones), would avoid artifact bias of the COG due to the exponential nature of few-electrons avalanche charge distribution.

$$\overrightarrow{COG_i} = \frac{\sum_{j=1}^{j=61} \vec{P_j} \cdot Q_{i,j}}{\sum_{j=1}^{j=61} Q_{i,j}} \tag{1}$$

In addition to the COG 2D histograms measured with given objects, others are measured without object (flat image). The ratio image is calculated according to equation 2, reflecting the transmission of the incident radiation through the object. The ratio image corrects for the inhomogeneity of the detector assembly. The background image is measured with no source.

$$Ratio = \frac{\left(\frac{Image_{object}}{Time_{object}} - \frac{Image_{background}}{Time_{background}}\right)}{\left(\frac{Image_{flat}}{Time_{flat}} - \frac{Image_{background}}{Time_{background}}\right)} \tag{2}$$

### 2.1.4   *Setup for point-like UV-photon imaging with the GPM*

Prior to cryogenic operation, the position resolution of the GPM with its pad readout was investigated at room temperature in a dedicated chamber outside of the LXe cryostat. This aimed at characterizing the dependence of the position resolution on the number of PEs due to statistical effect solely, avoiding the gamma-ray and neutron scattering effects within the LXe converter. The detector was illuminated through a fused-silica window by a point-like UV source ($H_2$ discharge lamp). The number of photons per pulse (from one to several thousands) was set using Oriel neutral density optical filters in front of the lamp. The number of photoelectrons per UV flash was derived from the pulse-height spectra recorded by the GPM. The trigger to the SRS electronics was provided by the electrical discharge pulse of the lamp.

### 2.1.5   *Setup for gamma-ray imaging with the LXe/GPM detector*

Two types of gamma-ray imaging experiments were performed with the full detector system: edge imaging with a broad beam irradiating a Pb object (covering half of the detector area) and detector irradiation with a collimated beam. Both experiments were performed with a disc-shaped $^{60}$Co source (1 mm-thick, $\varnothing$3.6 mm), emitting $3.7 \cdot 10^{7}$ $\gamma$/s of 1.17 and 1.33 MeV over $4\pi$. The detector was operated with Ne/5% CH$_4$ at gain of $4\times10^{4}$ at 647 mbar and 208 K (broad-beam experiments) and 475 mbar and 211 K (collimated-beam experiments). The flow was kept at 20 sccm throughout all measurements. In the broad-beam experiments a 12 mm-thick Pb plate, covering half of the detector's active area, was located under the cryostat (Figure 1A), 191 mm below the capillary-converter, 822 mm above the open (uncollimated) $^{60}$Co source. In the collimated-beam experiments the source was placed inside a Pb collimator (3 mm in diameter, 150 mm long), positioned on-axis 197 mm below the capillary-converter bottom. The calculated beam diameters at the capillaries bottom and top were ~7 mm and ~8 mm, respectively. The flat images (without object) for normalization were measured with an open source, located 1093 mm from the capillary-converter bottom.

### 2.1.6   *Setup for mixed neutron and gamma field imaging with the GPM/LXe detector*

As mentioned above, fast-neutron spectroscopy and neutron/gamma discrimination should be done by TOF. Experimental constraints (geometrical limitations at the laboratory and the low activity of the available AmBe source) prevented "neutron-only" imaging experiments. Hence, we performed Pb-edge imaging with the *mixed* neutron/gamma-ray field of the AmBe source.

A commercial encapsulated 96 mCi AmBe source, 10 mm in diameter and 10 mm length (active dimensions) was used, emitting ~2·10$^5$ neutrons/s into $4\pi$ in an energy range of 0-11 MeV. In addition it emitted ~1.4·10$^5$ 4.43 MeV $\gamma$/s over $4\pi$, via the $^9$Be($\alpha$, n; $\gamma$)$^{12}$C reaction; the source was wrapped with a thin Pb layer, absorbing the intense 59.5keV $^{241}$Am gamma line, and placed within a $\varnothing$25 mm bore in a paraffin-filled barrel ($\varnothing$380 mm, height=710 mm), 300 mm below the barrel's top. The barrel was covered by a 10 mm Pb lid, perforated at the center ($\varnothing$30 mm), absorbing gamma-rays emitted by neutron interactions within the barrel. The neutron-beam diameter, at the capillary-converter bottom, was ~60 mm. Imaging of the 12 mm-thick Pb edge was performed with the object located below the cryostat, 191 mm away from the capillaries bottom, and 500 mm above the collimated source. The GPM was operated at a gain of $2.4\times10^{4}$ under 475 mbar Ne/5%CH$_4$ at 212 K, under a flow of 20 sccm.

## 2.2 GEANT4 simulations

An extensive simulation study was performed using GEANT4 (version 9.3.2) [26], for predicting and validating the imaging performance under the specific experimental conditions described above (using an accurate 3D model of the cryostat and GPM). The standard GEANT4 model was used for gamma-rays and standard GEANT4 neutron high-precision model for neutrons with energies below 20 MeV. The calculations included all steps in the detection process, namely: gamma and neutron interaction probabilities; total deposited-energy distributions in LXe; total scintillation yields and their spatial distributions within the LXe volume; UV-photon transmission through the window; photon detection efficiency at the photocathode surface (including the QE and PE extraction and collection efficiencies); spatial distribution of the detected photoelectrons and their center of gravity (defining the position resolution of the detector).

In principle, the scintillation yield depends on the energy deposited in the LXe converter [27]. In the simulations we assumed the lowest scintillation yield values estimated by [27] for neutron-induced nuclear recoils and gamma-induced electron recoils: 4 photons/keV and 20 photons/keV, respectively. Furthermore, the GPM photon detection efficiency was set as $PDE_{GPM} = 10\%$ (at LXe emission wavelength of 175 nm). This $PDE_{GPM}$ is reasonable, given the effective quantum efficiency values measured for our CsI photocathode before installation (22%), and at the end of the experiments (8%) (see section 2.1.2).

Simulations were performed for all experiments mentioned above. In addition, infinitesimally thin pencil-beam simulations were performed for an ultimate detector projected for operational use (assuming $PDE_{GPM} = 20\%$). Such detector has a larger converter, of $580{\times}580{\times}50$ mm$^3$ and the GPM CsI photocathode is positioned at a shorter distance (15 mm rather than 34 mm) above the top of the capillaries. These latter simulations were performed for 1.3-15.1 MeV gamma-rays and 2-15 MeV neutrons relevant to the gamma/neutron radiography project (with $^{11}$B(d,n$\gamma$)$^{12}$C) discussed above [6, 10]. The simulations performed for the projected operational detector are a continuation of those discussed in detail in [12], with an updated capillary geometry consistent with the one used in the present experiments.

## 3 Results

### 3.1 UV-photon imaging with the GPM at room temperature

The position resolution of the GPM was determined at room temperature, with UV photons, prior to its installation in the LXe cryostat; the setup and methodology are described above. The detector was operated with Ne/5%CH$_4$; it was irradiated with a point-like source emitting UV-photon flashes, as described above. Spectra of the total UV-photon induced charge were recorded from all the GPM pads, event-by-event, for different PE yields per UV-flash; a charge threshold of 1.6 fC was set in each pad channel (Figure 5A). As expected, the charge spectra have Gaussian-like shapes for large number of PEs, while for few PEs the shape approaches a decaying exponent (due to avalanche buildup statistics).

Position profiles along the center of the 2D COG histograms, determined for the various number of PEs, are shown in Figure 5B. The position resolution (FWHM) values, calculated from these measured profiles, versus number of PEs, are shown in Figure 6, along with GEANT4 results. Note the good agreement between simulation and experiment.

The statistics of the number of PEs significantly affects the resolution for small numbers of PEs; in the present configuration it varied between sub-millimeter and a few centimeters.

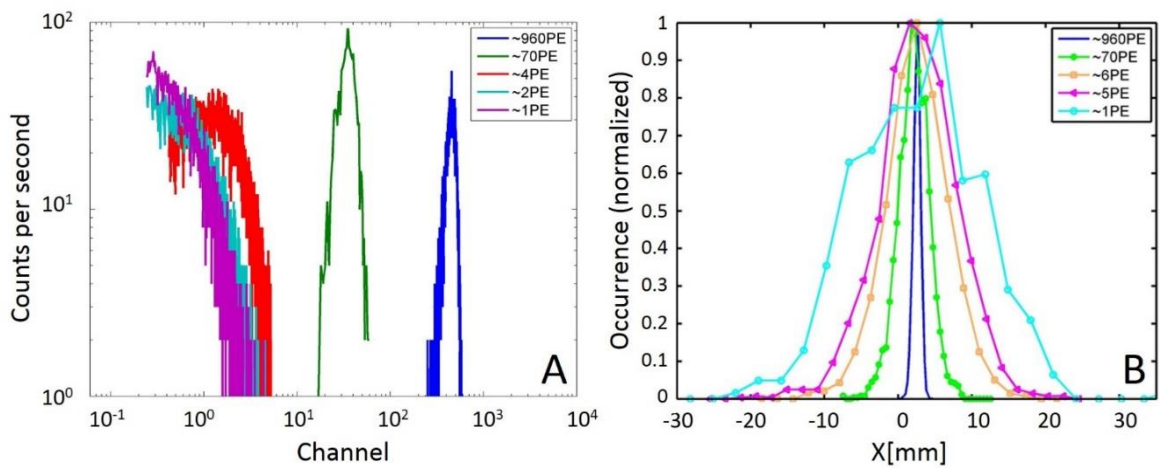

**Figure 5:** (A) Spectra of the total charge collected in all 61 pads of the readout electrode, for different numbers of UV-induced photoelectrons (note the log-log scale). (B) Profile along the x-axis of the COG histograms. Data measured with the triple-THGEM GPM (of Figure 2, without LXe convertor) for different numbers of photoelectrons per UV-lamp burst. Ne/5%CH$_4$; 1000 mbar, 298 K.

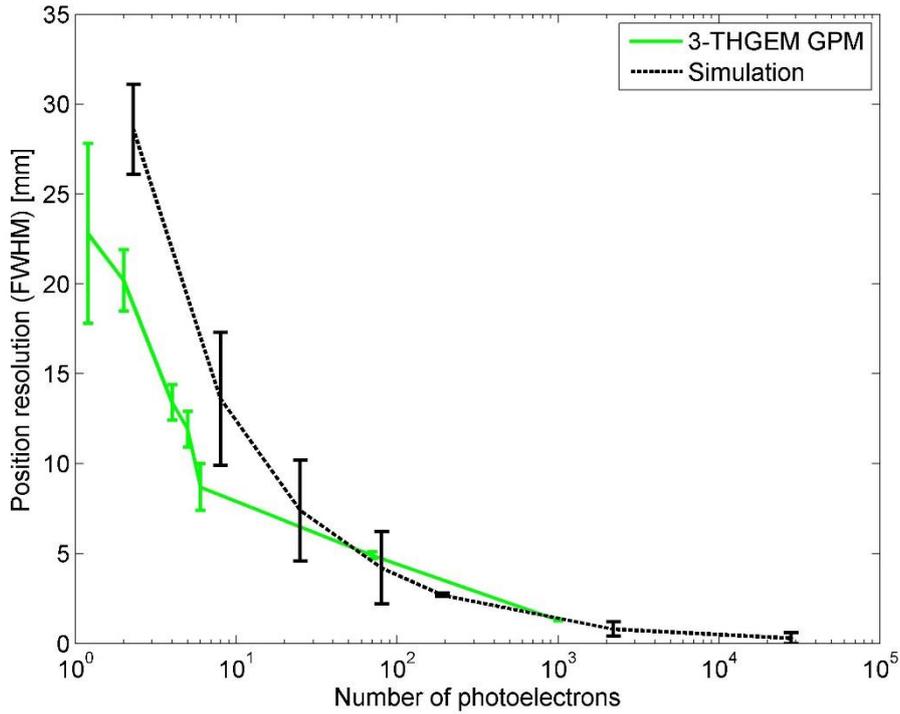

**Figure 6:** Summary of the measured position resolution derived from the distributions of Figure 5B, compared to the simulation results of the COG distributions, versus the number of photoelectrons. Ne/5%CH$_4$; 1000 mbar, 298 K.

## 3.2 Simulated PE and avalanche charge spectra for gamma-rays and neutrons

As a second interim step prior to operation under cryogenic conditions, we performed detailed simulations of the expected number of photoelectrons and avalanche charge distributions for the actual experimental setup. The simulations were done for a pencil beam irradiating the cryostat on-axis for the $^{60}$Co and AmBe source gamma and neutron energies (spectra for 15.1 MeV gammas were also included for reference). The results are shown in Figure 7. The average numbers of PEs are listed in Table 1.

Note that the exponential charge spectrum for the $^{60}$Co gamma-rays does not result from single-PE events, but rather reflects a large variance in the deposited energy and light collection from different interaction points inside the LXe converter, which is further smeared by avalanche statistics.

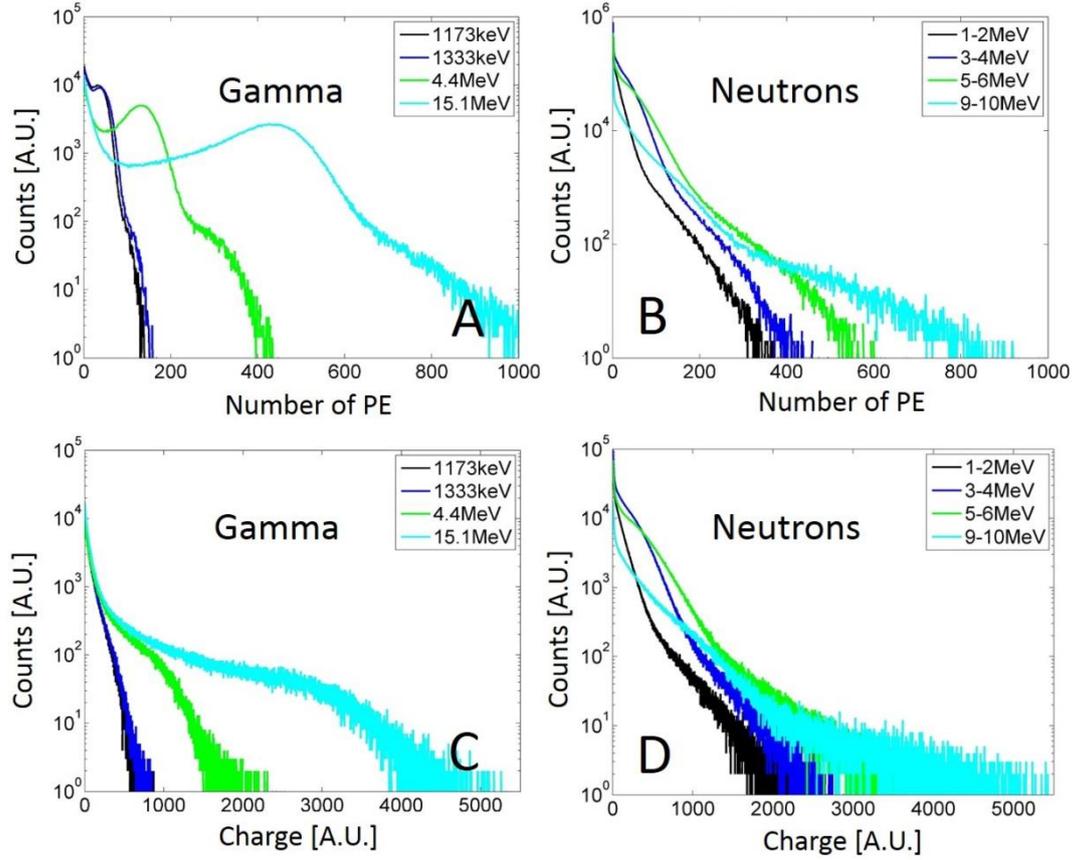

**Figure 7:** Gamma- and neutron-induced photoelectron distributions (A, B) and integrated-charge spectra (C, D), simulated for different gamma and neutron energies in the detector configuration of Figure 2. Detector $PDE_{GPM} = 10\%$, assumed gain $= 4.4 \times 10^4$ (see text).

**Table 1:** Calculated average number of PEs for the relevant gamma and neutron energies for the current experimental setup with GPM located 34 mm above the converter ($PDE_{GPM} = 10\%$).

| Gamma | | Neutrons | |
|---|---|---|---|
| Gamma Energy [MeV] | Average number of PEs | Neutron energy range [MeV] | Average number of PEs |
| 1.17-1.33 | ~30 | 0-4 | 19 |
| 4.4 | 98 | 4-7 | 36 |
| 15.1 | 297 | 7-11 | 44 |
| | | AmBe neutron spectrum | 32 |

### 3.3 Gamma-ray imaging with the LXe detector

The cryogenic gamma-ray imaging experiments were carried out in the setup described above (Figure 1 and section 2.1.5). Spectra of the total UV-photon charge, induced by gamma-interactions in the LXe converter/scintillator with and without the Pb object, were recorded event-by-event with different GPM gains (Figure 8). The spectra have exponential-like shapes, in agreement with the simulations (blue and black spectra in Figure 7C). As shown in Figure 8, the qualitative exponential-like shape of the spectrum does not depend on the gain.

For each event, the unweighted center of gravity was calculated using equation 1, setting the charge $Q_{ij}$ collected on all pads to 1 for all events (this is justified here, because the average number of PEs per pad is smaller than 1, as seen in Table 1); the resulting image of the Pb edge, calculated using equation 2, is shown in Figure 9A. The detector was operated at a gain of $3.8 \times 10^4$; the charge threshold was set to 1 fC and the pad threshold to 10 pads. The 2D image in Figure 9A represents the transmission of the incident radiation with the object, relative to the flat irradiation without object. One can clearly see the covered area of the detector.

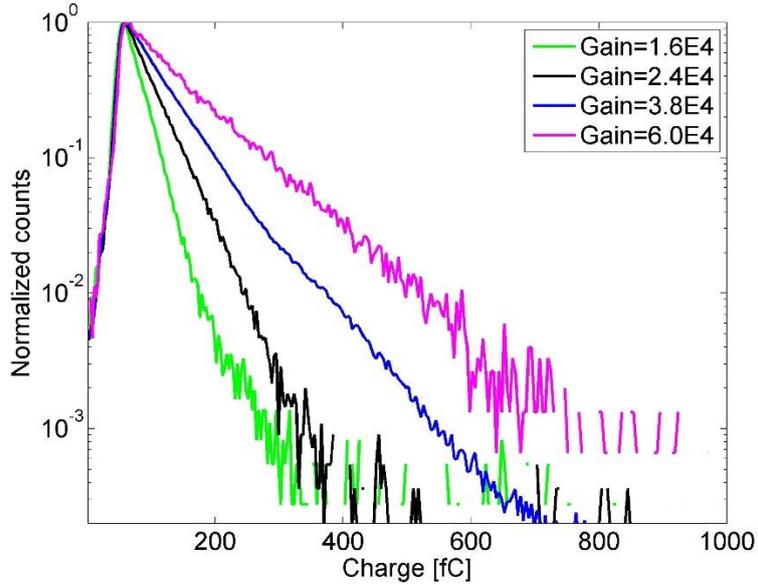

**Figure 8:** Gamma-ray interactions in the LXe converter (without the Pb object): spectra of the total UV-photon induced charge in all GPM pads, calculated offline event-by-event for different detector-gain values. No charge or pad thresholds were applied for the charge spectra.

The point spread function (PSF) of the imaging detector can be estimated by means of the edge spread function (ESF) technique [20, 28]. The ESF, shown in Figure 9B, is the average profile of the edge image. The expected transmitted fraction of $^{60}$Co gamma-rays through the 12 mm Pb object is 45%. In practice, the measured transmission was 55% (see Figure 9B), due to gamma scattering from the detector's uncovered area to the Pb-covered one and also by the object itself. This effect was validated by simulations. To smooth the statistical fluctuations, the edge profile was fitted with a logistic function (equation 3) [29], which models the ESF with adequate accuracy.

$$\text{ESF}(x) = a_0 + \frac{a_1}{1+\exp(-a_2(x-a_3))} \tag{3}$$

An approximate estimation of the PSF can obtained by differentiating the logistic function fitted to the ESF, as discussed below.

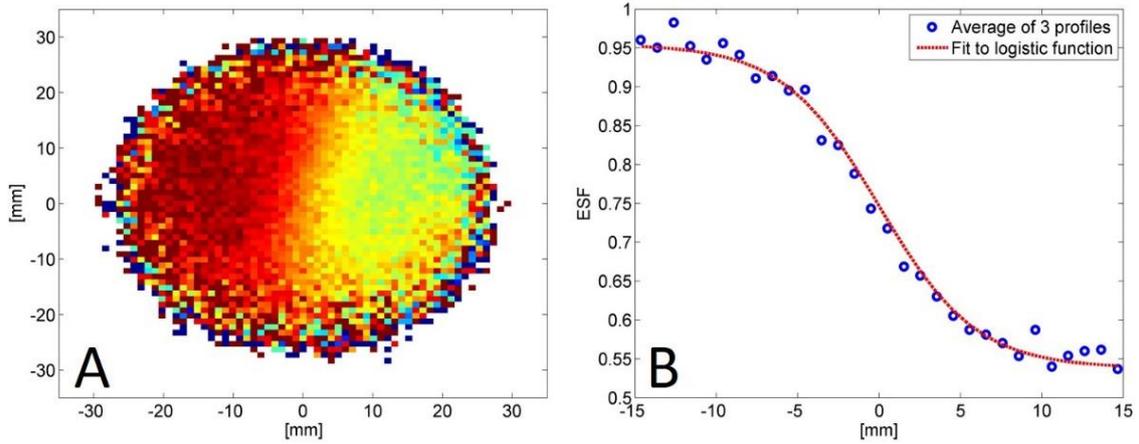

**Figure 9:** [60]Co gamma-ray image with the LXe/GPM detector (Figure 2), obtained by irradiating a lead object placed under the setup of Figure 1. (A) The 2D image of the object-to-flat ratio, calculated using equation 2; the color map indicates the transmission of the incident radiation through the object, relative to the case of no-object. (B) ESF Profile and fit to a logistic function (equation 3). GPM operating conditions: Ne/5%CH$_4$, 647 mbar, 208 K, GPM gain = $3.8\times10^4$.

Figure 10 shows the recorded image of the collimated [60]Co beam. The image represents the calculated ratio (equation 2) between an image taken with the collimator and one taken without it (*flat*). Note that the flat image was taken with the [60]Co source positioned at about 1 m below the LXe converter. The charge threshold was set to 1 fC and the pad threshold to 10 pads; the COG of each event was unweighted. The color map indicates the transmission of the incident radiation, relative to the flat one. The collimated beam image served for a second approximate estimation of the PSF.

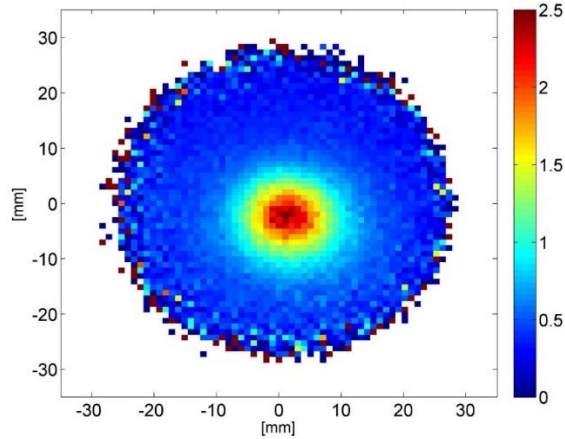

**Figure 10:** $^{60}$Co gamma-ray image obtained by irradiating the LXe/GPM detector (Figure 2) with a collimated-beam (~7 mm diameter at the converter level). The image shows the ratio of the beam-to-flat irradiation (unweighted COG-histogram-ratio image), as calculated according to equation 2. GPM operating conditions: Ne/5%CH$_4$, 475 mbar, 211 K, charge threshold = 1 fC, pad threshold = 10 pads, GPM gain = 4×10$^4$.

### 3.4 Edge imaging with the LXe detector in a mixed neutron & gamma field

The mixed-field irradiation setup with fast neutrons and 4.4 MeV gamma-rays and the procedures employed are described in section 2.1.6. As discussed above, the short distances involved in the present setup did not permit gamma-to-neutron separation by TOF. A typical spectrum recorded with the LXe/GPM detector is shown in Figure 11; it has an exponential shape with a long tail toward higher charges, as expected from the simulations (see Figure 7D). This is due to inelastic neutron collisions or neutron capture reactions, where the resulting gamma-rays can add their energy to the Xe recoil energy [12]. Furthermore, the introduction of hydrogen atoms inside the LXe (as a structure material of the Tefzel capillaries) may extend the neutron spectrum due to the contribution of knock-on protons, which may receive large fraction of the neutron energy in a single collision [12].

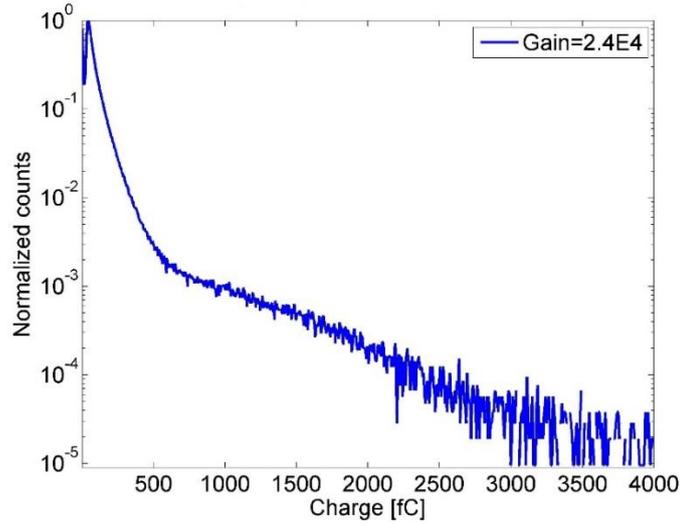

**Figure 11:** Mixed neutron and gamma-ray interactions in the LXe converter (without the Pb object): spectrum of the total UV-photon induced charge in all GPM pads, calculated offline event-by-event for a GPM gain of $2.4 \times 10^4$.

For each event, the unweighted center of gravity was calculated using equation 1, for all events. Similar to the $^{60}$Co gamma imaging, two measurements were done, with and without the Pb object. The charge threshold in both measurements was set to 1 fC and the pad threshold was set to 5 pads. The ratio image shown in Figure 12A was calculated using equation 2. The color map indicates the transmission of the incident radiation through the object, relative to the flat image. The average profile of the edge, the ESF, is shown in Figure 12B along with a fit to a logistic function (equation 3). One can clearly distinguish the covered area of the detector. The theoretical and simulated transmission of neutrons and gamma (of AmBe) through 12 mm Pb is 70%. In practice, the measured transmission was 82%, probably due to scattering from the concrete floor and walls, which was not taken into account in the simulations.

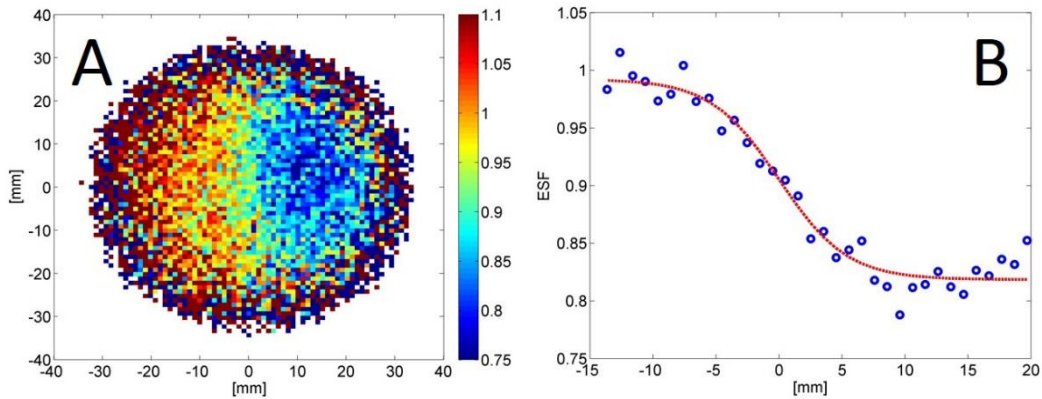

**Figure 12:** Lead object imaging with a mixed field of neutrons and gammas emitted from the AmBe source. (A) Ratio of the object to a flat image, calculated using equation 2. The color map indicates the transmission of the incidence radiation, relative to a flat image. (B)ESF profile and fit to a logistic function (equation 3). GPM operating conditions: Ne/5%CH$_4$, 475 mbar, 212 K, GPM gain = $2.4 \times 10^4$.

## 4    Position resolution analysis

### 4.1    Gamma imaging

For the Pb absorber edge irradiation measurements, an approximate estimation of the PSF was obtained for several pad-threshold values (Figure 13) by differentiating the average profile of the edge, i.e. the edge spread function (ESF) (Figure 9B). Table 2 compares the measured estimates and simulated results of the PSF FWHM in the present experimental geometry. The simulated values agree well with the experimental ones, validating the simulation tools. Note that "% of total counts" represents the detection efficiency of the events that interacted with the LXe converter, for a given pad threshold, i.e., the counting efficiency of converted events. The uncertainty on the measured position resolution values was estimated to be ~2 mm by fitting logistic function (equation 3) to a few ESF profiles, chosen from different regions on the object-to-flat ratio image (Figure 9A).

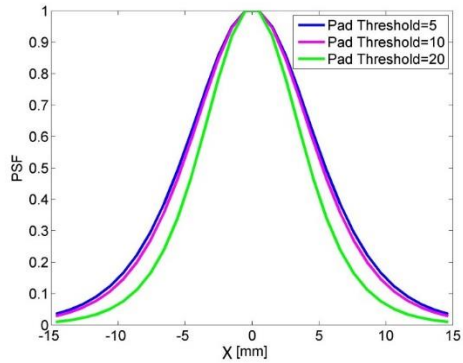

**Figure 13:** Approximate PSF of a [60]Co-irradiated Pb edge, for 5, 10 and 20 pad thresholds, as calculated by the derivative of the ESF of the Pb edge measurement (Figure 9B).

**Table 2:** Measured and simulated position resolution FWHM values resulting from the [60]Co-irradiated Pb edge, calculated for three pad thresholds. "% of total counts" represents the detection efficiency of interacting events for a given pads threshold (counting efficiency of converted events). The uncertainty on the measured spatial resolution values is ~2 mm.

| Pad threshold | % of total counts | PSF (FWHM) [mm] | |
|:---:|:---:|:---:|:---:|
| | | Measurement | Simulations |
| 5 | 99.1 | 12 | 12 |
| 10 | 71.6 | 11 | 11 |
| 20 | 14.3 | 9 | 10 |

Figure 14 shows the X and Y profiles through the center of the image of the collimated $^{60}$Co gamma-ray beam (7 mm diameter spot at the converter base) (Figure 10), fitted with a Lorentzian function. The FWHM of the measured profiles is again ~12 mm, similar to that of the PSF of Figure 13. The charge threshold was 1 fC and pad threshold was 10 pads. Table 3 lists the measured and simulated position resolution FWHM values. The uncertainty on the measured values was estimated as ~2 mm by the Lorentzian function fit.

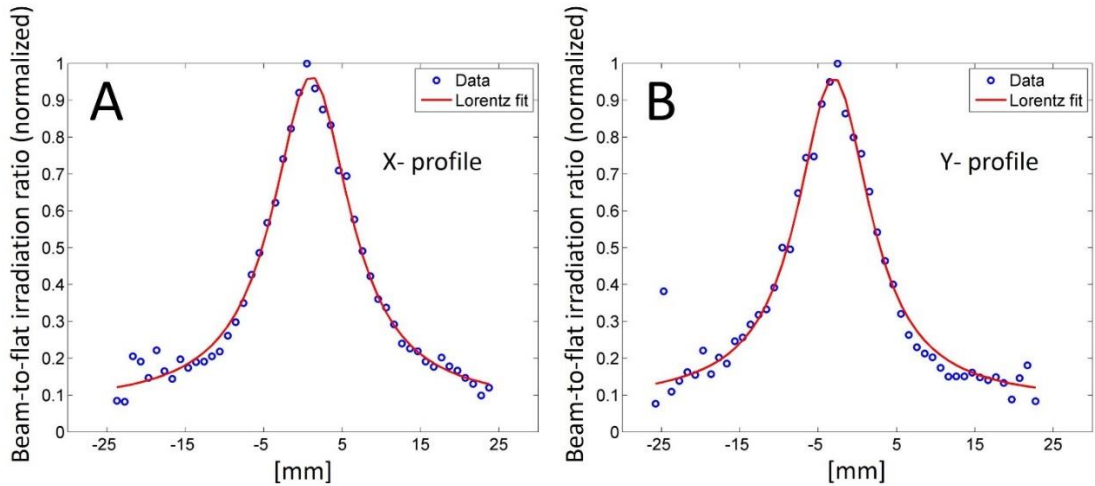

**Figure 14:** X and Y profiles through the center of the histogram of Figure 10 (collimated $^{60}$Co source), with a fit to a Lorentzian. GPM operating conditions: Ne/5%CH$_4$, 475 mbar, 211 K. Charge threshold = 1 fC, pad threshold = 10 pads, GPM gain = $4\times10^4$.

**Table 3:** Measured and simulated PSF values, calculated from the X and Y profiles of the collimated $^{60}$Co gamma beam, with unweighted COG. "% of total counts" represents the detection efficiency of interacting events for a given pad threshold (counting efficiency of converted events). The uncertainty on the PSF values is ~2 mm.

| Pad threshold | % of total counts | PSF (FWHM) [mm] | | |
| --- | --- | --- | --- | --- |
| | | Measurement | | Simulations |
| | | X profile | Y profile | |
| 5 | 95.8 | 12 | 12 | 10 |
| 10 | 63.1 | 12 | 11 | 10 |
| 20 | 14.4 | 11 | 11 | 9 |

In both edge-object and collimated-beam measurements with $^{60}$Co gamma-rays, the resulting spatial resolution (for a pad threshold of 5) was 12 ± 2 mm (FWHM) at high (>95%) counting efficiency of converted events; it is in good agreement with the simulated values for this detector configuration. While the PSF value did not improve significantly with increasing pad threshold from 5 to 10, the threshold change resulted in considerable loss of efficiency.

## 4.2    Mixed-field imaging

The PSF curves shown in Figure 15 were obtained by differentiating the logistic function fitted to the measured ESF (of Figure 12B), for pad thresholds of 5, 10 and 20 pads. Table 4 summarizes the measured and simulated position resolution values. The uncertainty on the measured spatial resolution values was estimated to be ~2 mm by fitting logistic function (equation 3) to a few ESF profiles, chosen from different regions on the object-to-flat ratio image (Figure 12A). Setting the pad threshold to 5 pads resulted in a reasonable counting efficiency of 89% of total counts (counting efficiency of converted events) and position resolution of $10 \pm 2$ mm (FWHM), in good agreement with simulations results (carried out for the present experimental conditions).

The slightly better spatial resolution obtained in the mixed-field experiment, compared to the [60]Co experiments, can be explained by the larger number of PEs for AmBe neutrons and 4.4 MeV gamma-rays (average PE numbers of 32 and 98 PE, respectively) compared to average of ~30 PE for [60]Co gamma-rays. Furthermore, the hydrogen content of the Tefzel capillaries improves the energy transfer by impinging neutrons.

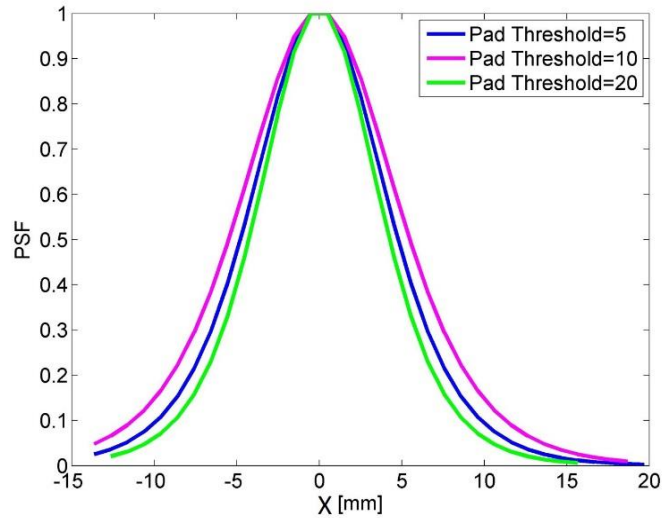

**Figure 15**: Mixed field of neutrons and gammas: PSF distributions for pad thresholds of 5, 10 and 20 pads, as calculated by the derivative of the ESF of the edge measurement (Figure 12B).

**Table 4:** Mixed field of neutrons and gammas: PSF values, calculated (with unweighted COG) from simulation and experimental edge-irradiation results, for various pads thresholds. "% of total counts" represents the detection efficiency of interacting events for a given pads threshold (counting efficiency of converted events). The estimated uncertainty on the PSF values is ~2 mm.

| Pad threshold | % of total counts | PSF (FWHM) [mm] | |
|---|---|---|---|
| | | Measurement | Simulations |
| 5 | 89.1 | 10 | 9 |
| 10 | 56.5 | 11 | 9 |
| 20 | 16.0 | 9 | 8 |

The experimental and simulation results with gamma rays and in a mixed radiation field (presented in Table 2-Table 4) are summarized in Figure 16.

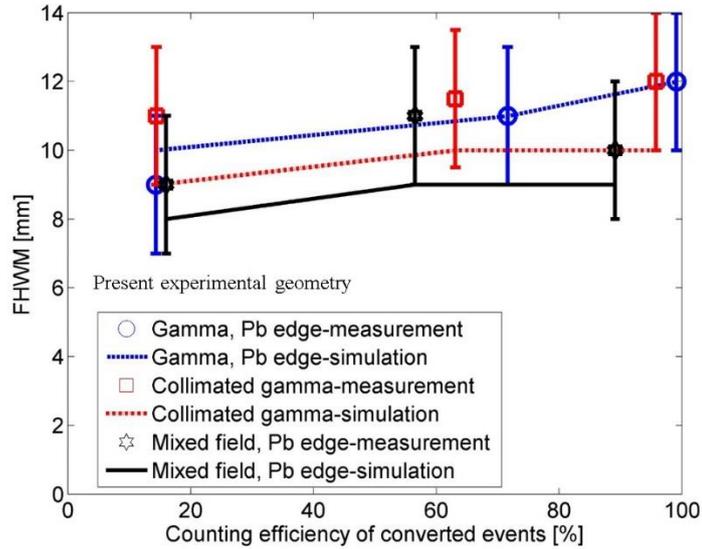

**Figure 16:** Measured and simulated position resolution values in the present experimental geometry versus detection efficiency of the events that interacted with the LXe converter. The counting efficiencies resulted from different pad-threshold values. The data are shown for gamma-ray imaging of a Pb edge object, a narrow gamma beam and edge imaging with mixed neutron & gamma field.

## 4.3    Projected position resolution in a large-area operational detector

The rather large PSF values of ~9-12 mm (FWHM) obtained in the present setup for gamma-rays and neutrons were due to its un-optimized design, namely the large distance (34 mm) between the 100 mm-diameter photocathode and the top edge of the capillary array. While this distance should not play an important role in a large-area detector (i.e., a detector with lateral dimensions much larger than this distance), it had a major effect in the present study. Because of the relatively small diameter of the GPM, a large fraction of the photons emitted from the top of the capillaries was refracted out of the sensitive area of the detector, thus severely degrading its position resolution. A simulation of the present setup, in which the photocathode was brought to a distance of ~3 mm from the window, resulted in a major improvement in the calculated FWHM for $^{60}$Co MeV gamma-rays: ~2 mm compared to ~12 mm.

The good agreement between the measured and simulated position resolution values in the present setup indicates that all of the important experimental factors were accounted for, and therefore validates the simulation tools. Simulation studies performed with the same set of validated tools predict a much better position resolution for a projected operational system comprising large-area LXe converter and GPM irradiated with pencil beams of both gamma-rays and neutrons. The simulated setup comprised, in this case, a LXe converter with Tefzel capillaries with an overall size of 580×580×50 mm$^3$ coupled to a GPM with a similar area, with a CsI photocathode 15 mm away from the LXe converter edge (5 mm behind a 10 mm-thick fused silica window). The GPM PDE was assumed to be 20%. The setup was irradiated perpendicularly at its center with a pencil beam of either gamma rays (1.3–15.1 MeV) or neutrons (2–15 MeV). The results (detection efficiency and position resolution) are shown in Figure 17 for different gamma and neutron energies as a function of varying threshold values of the number of photoelectrons. For neutrons, the position resolution is better than ~2 mm (FWHM) at all energies with a corresponding detection efficiency of ~20% (Figure 17A and B). For gamma-rays, the position resolution is 2.4 and 3.5 mm (FWHM) at 4.4 and 15.1 MeV respectively, with a constant detection efficiency of ~30% (Figure 17C and D). The increase in FWHM with gamma energy results from increased scattering. The weak dependence of both the detection efficiency and position resolution on the set threshold is due to the large numbers of photoelectrons (90-990 PEs for 1.3-15.1 MeV gammas, 70-340 PEs for 2-15 MeV neutrons). Additional simulations showed that, because of the large number of PEs, relaxing the assumption of 20% PDE to 10% has a small effect on the detector's performance for the considered neutron and gamma energy ranges.

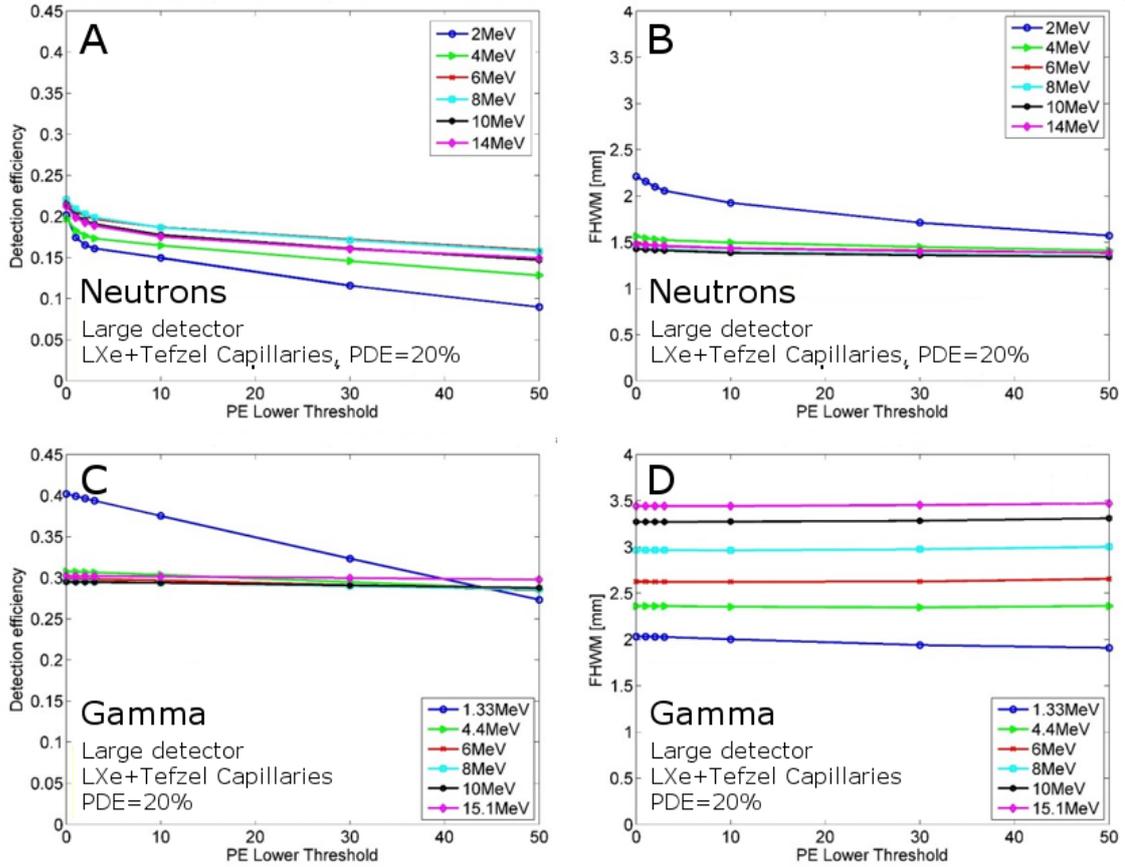

**Figure 17:** Simulated detection efficiencies (left side) and position resolutions (right side) vs. threshold on the number of photoelectrons in each event, for pencil beams of neutrons (A, B) and gamma-rays (C, D), irradiating a large-area detector (LXe within Tefzel capillaries array of 580×580×50 mm³, $PDE_{GPM}$ =20%).

## 5 Summary and conclusions

The imaging properties of a 100 mm diameter (active-area) detector prototype, conceived for simultaneous imaging and spectroscopy of fast neutrons and gamma-rays, are described. The detector comprises a LXe converter/scintillator contained in "fiber-like" Tefzel capillaries, coupled to a UV-sensitive imaging gaseous photomultiplier.

Imaging experiments with [60]Co low-energy gamma-rays and with a mixed fast-neutron/gamma radiation field of an AmBe source were performed at the laboratory. In the present experimental configuration, [60]Co gamma rays yielded spatial resolutions with a PSF of 12 ± 2 mm (FWHM), derived from edge irradiation of a Pb object and of a collimated gamma-ray beam. The results, with counting efficiency of converted events >95%, are in good agreement with simulation results.

Similar edge-imaging experiments with the mixed gamma (4.4 MeV) and neutron (0-11 MeV) radiation field (ratio of gamma-to-total-neutrons $R_{\gamma/n}$=0.575) were performed, yielding PSF localization resolutions of ~10 ± 2 mm (FWHM), under counting efficiency of converted events of ~90%; they are also in good agreement with simulations results.

Simulations for estimating the position resolution values of a large-area detector were performed with the same set of validated simulation tools for pencil beams of 2-15 MeV neutrons (relevant to FNNR) and for 15.1 and 4.4 MeV gamma-ray energies (relevant to DDEG). Respective fast-neutron and gamma position resolutions ~2 mm and 2.5–3.5 mm (FWHM), were obtained for a GPM with its photocathode located 15 mm above the top surface of the converter ($580 \times 580 \times 50$ mm$^3$ LXe-filled Tefzel-capillaries); the GPM had a $PDE_{GPM} = 20\%$, but due to the large numbers of photons rather similar results were obtained also with $PDE_{GPM} = 10\%$.

Note that the assumption of 20% PDE was taken for photons which have already passed through the window; i.e., it represents the probability that a photon transmitted through the window gives rise to a detectable photoelectron. This value should be regarded as an idealized upper bound; it may be reached, for example, if one assumes a QE of 25% at 175 nm (the reference value of CERN-RD26 [30]), an overall extraction efficiency of 86% (which may, in principle, be possible in highly quenched mixtures such as Ne/50%CH$_4$ [23]), a 98% transparent mesh and 95% probability to generate a signal above the electronic noise. Reaching a PDE of 10% for photons passing the window, on the other hand, can be readily achieved by greatly relaxing all of these parameters.

Neutron energy resolution of ~500 keV at 8 MeV is required for resolving neutron resonances of interest in FNNR. For a 6 m long TOF facility, this energy resolution would be equivalent to a time resolution of ~5 ns. The time resolution of the GPM depends on THGEM geometry and on the number of PEs per pulse. Preliminary TOF results, performed with AmBe 4.4 MeV gamma-rays under the present experimental constrains (mainly geometrical limitations and low AmBe source intensity), resulted in time resolutions of ~10 ns (FWHM) (for a GPM gain of $7.6 \times 10^4$). According to the prototype simulations described above (see section 2.2), the average number of PEs resulting from 4.4 MeV gamma-ray interactions is 98 (see Table 1). However, the expected time resolution for this number of PEs should be ~2 ns (FWHM), based on previous studies [21, 31, 32]. According to these studies and to Table 1, the expected time resolution for 0-11 MeV neutrons (average PE number of 19 - 44 PEs, respectively) would be 5.5 – 3 ns.

Improving the time resolution will be the subject of further studies. One possible way would be to adopt different multiplier geometries, with faster avalanche process and shorter photoelectron collection times from the CsI-coated electrode. Further improvement in timing could be reached by using faster counting gases with reduced electron diffusion parameters – keeping however high PE extraction efficiency from the photocathode (preserving high PDE of the GPM). Possible candidate could be Ne/20%CH$_4$, or even Ne/50%CH$_4$ [23]; compared to the presently employed Ne/5%CH$_4$, this will result in more favorable electron-transport parameters [33].

The results and the estimates made above indicate that the detector concept investigated in this work has the potential of offering a cost-effective, large-area solution for the **simultaneous** efficient imaging of fast-neutrons and gamma-rays, with moderate spatial resolutions. The detector investigations have been carried out so far at the laboratory, with radioactive sources. Simulation tools, validated by the experimental results, permitted indicating that this detector concept has the potential of fulfilling the efficiency and resolution requirements [9] of a system for inspecting large objects, such as marine and aviation containers (e.g. efficiency >10% for neutrons and gamma-rays and spatial resolution in the order of 5-10 mm FWHM). Naturally, further studies of the required neutron and gamma-ray fields are necessary, with optimized GPM detector configurations and electronics, to fully validate the concept for the above applications.

Among leading competing detection techniques, are for example the TRION (neutrons spectroscopy) [34] and TRECOR (gamma and neutrons spectroscopy) [35] systems which combine solid-scintillator screens and intensified CCD cameras. Compared to the proposed LXe detector, they possess similar position resolutions (~1 mm FWHM) for neutrons and gamma-rays and better timing (affecting energy resolution of neutrons). However they require integration of the gamma and neutron images, which are measured separately in different detector media. Furthermore, the very high cost of large-area imagers of this type, required for an operational container screening system, could be exorbitant.

**Acknowledgments**

The authors would like to thank Dvir Hamawi, Amil Malka, Sharon Hliva, Zion Versolker, Moshe Kendelker, Yitzhak Halamish, Nissim Mimon, Hanoch Apteker and Sharon Biton from NRCN for their appreciated assistance in preparation of the collimators for the neutron and gamma sources. This work was partly supported by the PAZY Foundation (Grant No. 258/14), by the Minerva Foundation with funding from the German Ministry for Education and Research (Grant No. 710827) and by the Israel Science Foundation (Grant No. 477/10).